\documentclass[lettersize,journal]{IEEEtran}

\usepackage{amsmath,amsfonts}
\usepackage{bm}
\usepackage{algorithmic}
\usepackage{algorithm}
\usepackage{array}
\usepackage[caption=false,font=normalsize,labelfont=sf,textfont=sf]{subfig}
\usepackage{textcomp}
\usepackage{stfloats}
\usepackage{url}
\usepackage{verbatim}
\usepackage{graphicx}
\usepackage{xcolor}
\usepackage{soul}
\usepackage{ulem}
\usepackage{cite}

\begin{document}

\renewcommand{\baselinestretch}{0.995}

\title{Robust Synchronous Reference Frame Phase-Locked Loop (PLL) with Feed-Forward Frequency Estimation}

\author{Michael Ruderman, Elia Brescia, Paolo Roberto Massenio, Giuseppe Leonardo Cascella, David Naso
\thanks{M. Ruderman is with the Department of Engineering Sciences, University of Agder, 4879 Grimstad, Norway (e-mail: michael.ruderman@uia.no). He is on annual sabbatical at Polytechnic University of Bari. }
\thanks{E. Brescia, P. R. Massenio, G.L. Cascella, D. Naso are with the Department of Electrical
and Information Engineering, Polytechnic University of Bari, Bari, Italy (e-mail:
elia.brescia@poliba.it; paoloroberto.massenio@.poliba.it; giuseppeleonardo.cascella@poliba.it; david.naso@poliba.it).
}

\thanks{\textcolor[rgb]{0.00,0.00,1.00}{Author's manuscript accepted in IEEE Transactions on Control Systems Technology (TCST), May 2026} 
}

}



\maketitle

\begin{abstract}
Synchronous reference frame phase-locked loop (SRF-PLL) techniques are widely used for interfacing and control applications in the power systems and energy conversion at large. Since a PLL system synchronizes its output with an exogenous harmonic signal, often 3-phases voltage or current, the locking of the frequency and phase angle depends on the performance of the feedback loop with at least two integrator terms, and on the distortions of the measured input quantities. For the conventional SRF-PLL with a proportional-integral (PI) control in feedback, we are providing a robust design which maximizes the phase margin and uses the normalization scheme for yielding the loop insensitive to the input amplitude variations. The main improvement in the transient behavior and also in tracking of frequency ramps is achieved by using the robust feed-forward frequency estimator, which is model-free and suitable for the noisy and time-varying harmonic signals. The proposed feed-forward-feedback SRF-PLL scheme is experimentally evaluated on the 3-phases harmonic currents from a standard PMSM drive with the varying angular speeds and loads. Both, the tracked angular frequency and locked phase angle are assessed as performance indicators of the proposed SRF-PLL with feedforwarding.         
\end{abstract}


\newtheorem{theorem}{Theorem}
\newtheorem{remark}{Remark}

\section{Introduction}
\label{sec:1}

Phase locked loops (PLL) systems are essential for synchronization and control in power electronics, sensorless motor drives, and interfacing of energy conversion systems. PLLs date back to the first half of the twentieth century when first designed and used for the synchronous reception of radio signals \cite{DeBell1932}. Since then, PLL became an essential tool in different engineering fields, here we refer solely to some standard textbooks e.g. \cite{margaris2004,egan2007}, this for the sake of brevity and basics. A comprehensive practical reference on PLL techniques can also be found in \cite{gardner2005}. At large, PLL constitutes a circuit (or digital implementation) that synchronizes the signal of the oscillator with the second input signal, also denoted as reference, so that both operate at the same frequency and have a desired phase shift (oftentimes zero phase shift). Extensive reviews on the three-phase PLLs and their applications (like the one for sensorless motor control) can be found e.g. in \cite{golestan2017} and \cite{wang2021}, respectively. The rigorous mathematical definitions and control theoretical problems of PLL circuits are provided in \cite{leonov2015}. 

Among PLL techniques, the synchronous reference frame PLL (further as SRF-PLL) is the most widely used in three phase applications due to its simplicity and effectiveness. It typically consists of a phase detector (PD), proportional integral controller as a loop filter (LF), and a voltage controlled oscillator (VCO), cf. \cite{margaris2004,karimi2004,egan2007}. PLLs can be classified by their type based on the number of integrators in the loop. A type-1 PLL can track the phase steps but not the frequency steps, while a type-2 PLL can eliminate steady-state error for step frequency changes but fails under ramp frequency inputs. Recall that the latter are common in servo-systems such as motor drives \cite{wang2021,novak2022}. High loop gains can partially compensate for the frequency tracking errors but reduce the phase margin and increase noise sensitivity \cite{mathew2024,wu2019}. As a significant motivating example, the wind turbine converters can experience loss of synchronisation from the grid fundamental frequency \cite{goksu2014}, so that the PLL robust stability and performance play an essential role. While higher-order PLLs (e.g., type-3 PLL) theoretically eliminate ramp errors, they compromise robustness by reducing the phase margin and increasing the noise sensitivity \cite{wang2021}. \textcolor[rgb]{0.00,0.00,0.00}{Note that the robustness of PLL-based converters' connection, given the situation of different grid disturbances, is one of the key problems for successful integration of renewable energy sources into the grid}. For instance, a PLL-based anti-islanding methods was discussed in \cite{ciobotaru2010} and a boundedness of the phase and frequency estimation errors in the presence of disturbances was recently addressed e.g. in \cite{mathew2024}. 

One of the possible approaches to increase the PLL performance involves a feedforward injecting the known (or accurately estimated) frequency signal into the PLL loop, cf. the control diagram in the seminal work \cite[Fig.~1]{kaura1997}, also either after the PI controller or after its integrator \cite{karimi2012}. Recall that when the feedforward estimate provides a nearly constantly changing frequency, the equivalent input becomes a step signal, thus enabling ramp tracking without increasing the loop order, cf. \cite{golestan2017}. \textcolor[rgb]{0.00,0.00,0.00}{This fact serves as one of the main motivations for our approach without changing relative degree of the PLL loop transfer function.} Several strategies were proposed to generate the feedforward term. Injecting the given nominal frequency works under steady-state conditions but fails in dynamic or unknown operation modes \cite{wu2019,kaura1997}. More adaptive approaches are based on frequency locked loops (FLLs) or second-order generalized integrators (SOGIs) to provide the feedforward signal. But both are sensitive to the offsets and harmonics, and require additional loops and tuning parameters cf. \cite{wang2021}. Also worth noting is that the observer-based schemes can provide a better accuracy in some sensorless systems but depend largely on the system parameters and suffer from the modeling errors \cite{Bierhoff2017,Lascu2020}.

This paper proposes a novel SRF-PLL architecture enhanced by a robust and model-free frequency estimator, introduced in \cite{ruderman2022}, used to generate the feedforward signal. \textcolor[rgb]{0.00,0.00,0.00}{In contrast to several existing techniques, the proposed estimator operates directly on the measured three-phase signals, thus without any need to model the underlying dynamic system or to introduce the additional filtering loops or observers. The proposed method  ensures robustness to the amplitude and frequency variations and, especially, to the measurement noise, and it requires only one tuning parameter in addition to a separate design of the conventional SRF-PLL. Injecting the robust online frequency estimate into the SRF-PLL loop (see the overall block diagram shown later in Fig. \ref{fig:OverallScheme}) allows for an accurate frequency and phase locking even in presence of the load transients and frequency variations. Remarkable feature of the proposed approach is that this is accomplished without increasing the loop order and, thus, without  compromising its robustness. Unlike the other higher-order PLL techniques, i.e. with loop relative degree higher than two, our approach keeps two integrators in the loop, as provided in the seminal work \cite{kaura1997} and still used in the most of industrially operational PLL implementations. The proposed implementation of SRF-PLL with feed-forward (denoted as SRF-PLL-FF) is real-time executable and does not require any extensive pre- or post-processing of the signals.} We also propose a power-invariant normalization scheme, in order to decouple the loop gain from the input amplitude, thus improving further stability and simplicity of the PLL tuning. Experimental evaluations are based on the PMSM drive phase currents, under dynamic speed and torque variations, and confirm that the SRF-PLL-FF scheme significantly outperforms the standard SRF-PLL one.

The remainder of the paper is as follows. Section II revisits the SRF-PLL structure with focus on its small-signal modeling and symmetrical optimum loop design. Section III provides the robust frequency estimator used to generate the feedforward signal. Section IV presents the experimental study on a PMSM drive system. Concluding remarks are in Section V.

\section{Synchronous reference frame phase-locked loop system}
\label{sec:2}

\subsection{PLL principle}
\label{sec:2:sub:1}

\begin{figure}[!h]
\centering
\includegraphics[width=0.85\columnwidth]{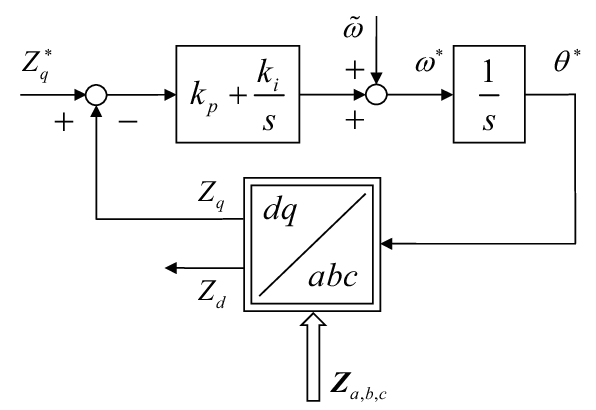}
\caption{Block diagram of conventional SRF-PLL.}
\label{fig:PLL}
\end{figure}

The conventional SRF-PLL structure is shown in Fig. \ref{fig:PLL}, cf. \cite{kaura1997}, where the PI regulator described by the $k_p + k_i/s$ transfer characteristics (with the complex Laplace domain variable $s$) adjusts a virtual frequency quantity $\omega^{\ast}$. The integration of the latter yields the phase angle $\theta^{\ast}$. This aims to be synchronized with the phase angle $\theta$ of the three-phase input $\bm{Z}_{\{a,b,c\}}$, usually the alternating voltage or current quantities. The control parameters of a conventional SRF-PLL are $k_p,k_i > 0$, while $\tilde{\omega}$ denotes the nominal frequency entering the synchronization loop in a feed-forward manner. Worth noting is that the knowledge of the nominal (or effective) $\tilde{\omega}$-value drives a PLL system closer to the $(\theta,\omega)$ operation point and, thus, facilitates the function of regulator to a large extent \cite{kaura1997}.

Assuming the three-phase input signals\footnote{Frequently, the input voltage signals are assumed in context of a PLL-based synchronization. Notwithstanding, also the current or rotating flux or back electromotive force signals can appear as either three- or two-phase input quantities $\bm{Z}_{n}$ used by SRF-PLL. The present study focuses on the sampled $n \in \{a,b,c\}$ currents of an inverter-operated PMSM drive.} 
\begin{equation}
\bm{Z}_{\{a,b,c\}} = Z \, \left(
                            \begin{array}{c}
                              \cos\bigl(\theta\bigr) \\[2mm]
                              \cos\Bigl(\theta - \dfrac{2\pi}{3}\Bigr) \\[2mm]
                              \cos\Bigl(\theta - \dfrac{4\pi}{3}\Bigr) 
                            \end{array}
                          \right),
\label{eq:2:1}
\end{equation}
with $Z$ to be the magnitude (which can be often unknown or time-varying) of the three-phase electric signals, the $abc$-$dq$ nonlinear transfer block provides the standard Clarke's and Park's transformations \cite{clarke1943,park1929}. Following to that, the output quantities of the rotating reference frame  are given by  
\begin{equation}
\left(
  \begin{array}{c}
    Z_d \\
    Z_q \\
  \end{array}
\right) =  T \times \bm{Z}_{\{a,b,c\}},
\label{eq:2:2}
\end{equation}
where 
\begin{equation*}
T = \frac{2}{3} \left[
                                       \begin{array}{ccc}
                                       \cos\bigl(\theta^{\ast}\bigr) & \cos\Bigl(\theta^{\ast}-\dfrac{2\pi}{3}\Bigr) & \cos\Bigl(\theta^{\ast}-\dfrac{4\pi}{3}\Bigr) \\[3mm]
                                       -\sin\bigl(\theta^{\ast}\bigr) & -\sin\Bigl(\theta^{\ast}-\dfrac{2\pi}{3}\Bigr) & -\sin\Bigl(\theta^{\ast}-\dfrac{4\pi}{3}\Bigr)
                                       \end{array}
                                     \right].
\end{equation*}
Substituting \eqref{eq:2:1} into \eqref{eq:2:2} results in 
\begin{equation}
\left(
  \begin{array}{c}
  Z_d \\
  Z_q \\
  \end{array}
\right) = Z \left(
              \begin{array}{c}
              \cos\bigl(\theta^{\ast} - \theta \bigr) \\[1mm]
              \sin\bigl(\theta^{\ast} - \theta \bigr) 
              \end{array}
            \right).
\label{eq:2:3}
\end{equation}
It can be seen that both quantities of the rotating reference frame depend trigonometrically on the synchronization error $\Delta \theta = \theta^{\ast} - \theta$. If the synchronization error is small, i.e. SRF-PLL remains in vicinity to the operation point $\Delta \theta \approx 0$, both trigonometric nonlinearities lead to $Z_d \approx Z$ and $Z_q \approx 0$. This forms the basis for a simplified (often denoted as small-signal) modeling of the SRF-PLL systems, cf. \cite{kaura1997,golestan2017}, that is suitable for a feedback control analysis and design (cf. section \ref{sec:2:sub:2}). 

\textcolor[rgb]{0.00,0.00,0.00}{Despite the SRF-PLL output quantity $Z_d$ provides a measure of the amplitude of the three-phase signal and is, therefore, often used for an amplitude normalization scheme within a SRF-PLL structure, see \cite{golestan2017}, its inherent high-frequency distortions require an additional low-pass filtering. This type of amplitude normalization is, however, one of the most used in PLL practice. Another magnitude normalization approaches operate already in the stationary frame (i.e. $dq$-frame) like for example the one proposed as a complex-type normalization extension, see \cite{dai2024complex} and references therein.} In order to bypass a phase lag of $Z_d(t) \approx Z$ associated with it, we provide an alternative scheme of the amplitude normalization which is based on the fact that the $abc$-$dq$ transformation is power-invariant. By introducing the normalization factor
\begin{equation}
N \bigl( \bm{Z}_{\{a,b,c\}} \bigr) = \sqrt{Z_a^2(t) + Z_b^2(t) + Z_c^2(t)}
\label{eq:2:4}
\end{equation}
of the rotary vector, one obtains   
\begin{equation}
\bar{\bm{Z}}_{\{a,b,c\}} = N^{-1}\bigl( \bm{Z}_{\{a,b,c\}} \bigr) \times \bm{Z}_{\{a,b,c\}}. 
\label{eq:2:5}
\end{equation}
It can easily be shown that independently of the $Z$-magnitude of the $\bm{Z}_{\{a,b,c\}}$ input, the magnitude of the normalized three-phase signal \eqref{eq:2:5} remains always constant as
\begin{equation}
\bar{Z} = \max \bar{\bm{Z}}_{n} = \sqrt{\frac{2}{3}}, \quad \forall \; n \in \{a, b, c\}.
\label{eq:2:6}
\end{equation}
Following to that, the amplitude-normalized signal \eqref{eq:2:5} is used as the SRF-PLL input instead of $\bm{Z}_{\{a,b,c\}}$, cf. Fig. \ref{fig:PLL}, so that the constant magnitude \eqref{eq:2:6} is further assumed instead of $Z$.

\subsection{Open loop small-signal analysis}
\label{sec:2:sub:2}

For small values of $\Delta \theta$, the trigonometric term in feedback behaves linearly, i.e. $\sin(\Delta \theta) \approx \Delta \theta$, so that the SRF-PLL can be (locally) considered as a linear control system, cf. \cite{kaura1997,golestan2017,wang2021}. Note that assigning $Z_q^{\ast}$ to zero, cf. Fig. \ref{fig:PLL}, allows directly to lock to the frequency and phase of the input signal by regulating $Z_q(t) \rightarrow 0$ and so $\Delta \theta \rightarrow 0$. Following to that, the open-loop transfer function (see e.g. \cite{maciejowsk1989} for basics) of the SRF-PLL with the normalized amplitude \eqref{eq:2:6} becomes 
\begin{equation}
H(s) = \sqrt{\frac{2}{3}} \, \frac{k_p s + k_i}{s^2} \frac{1}{\tau s + 1}.
\label{eq:2:7}
\end{equation} 
Note that the last term of the loop transfer function \eqref{eq:2:7}, which is not included into the block diagram in Fig. \ref{fig:PLL}, takes into account an inherent sampling delay. The latter is approximated by the first-order transfer characteristics with unity gain, see \eqref{eq:2:7}, while the sampling time satisfies $0 < \tau \ll k_p / k_i$.

Since the open-loop transfer function \eqref{eq:2:7} has the phase angle value in the output channel, another associated frequency open-loop transfer function, cf. Fig. \ref{fig:PLL}, is  
\begin{equation}
H_{\omega}(s) = \sqrt{\frac{2}{3}} \, \frac{k_p s + k_i} {\tau s + 1} \frac{1}{s} \equiv G_{c}(s) \frac{1}{s}.
\label{eq:2:8}
\end{equation} 
Here $G_{c}(s)$ summarizes the transfer elements which are belonging to the sampled control system, this for the sake of brevity in the below following analysis. Introducing the frequency error transfer function $E_{\omega}(s) = \bigl( \omega - \omega^{\ast}\bigr)/\omega $ of the SRF-PLL feedback loop, one obtains 
\begin{equation}
E_{\omega}(s) = \frac{1}{1 + H_{\omega}(s)} = \frac{s}{s + G_{c}(s)}.
\label{eq:2:9}
\end{equation} 
If the inputs have the frequency ramp changes, given by $\kappa / s^2$ where $\kappa > 0$ is the ramp inclination factor, one can easily show by applying the final value theorem (see e.g. \cite{antsaklis2007}) that the steady-state frequency error $\Delta \omega = \omega^{\ast} - \omega$ results in 
\begin{equation}
\Delta \omega (s) \bigr|_{s \rightarrow 0} =  \underset{s \rightarrow 0} {\lim} \: s  E_{\omega}(s) \frac{\kappa}{s^2}  = \sqrt{\frac{3}{2}} \, \frac{\kappa}{k_i} \neq 0.
\label{eq:2:10}
\end{equation} 
This implies that independently of the design of the PI control with $0 < k_i < \infty$, the SRF-PLL scheme has always a non-zero frequency estimation error in case the input signal $\bm{Z}_{\{a,b,c\}}$ is subject to a frequency ramp, cf. e.g. \cite{wang2021}. Recall that the analysis shown above is valid for linearization around the operation point $\Delta \theta \approx 0$ only, while an exact steady-state error can equally be analyzed by incorporating the corresponding trigonometric functions, cf. \cite{leonov2015,golestan2017}. It becomes also evident that if a feed-forward injection of the frequency estimate $\tilde{\omega}$ (as in Fig. \ref{fig:PLL}) provides a nearly constant frequency change within the SRF-PLL control loop (that is equivalent to the steady-state estimation error $\tilde{\omega}-\omega$), then the equivalent input will contain a single integrator, i.e. $\texttt{const} / s$. Therefore, by using the same argumentation and final value theorem, one can show that the steady-state frequency estimation error tends to zero.

\subsection{Symmetrical optimum phase margin maximization}
\label{sec:2:sub:3}

\begin{figure}[!h]
\centering
\includegraphics[width=0.98\columnwidth]{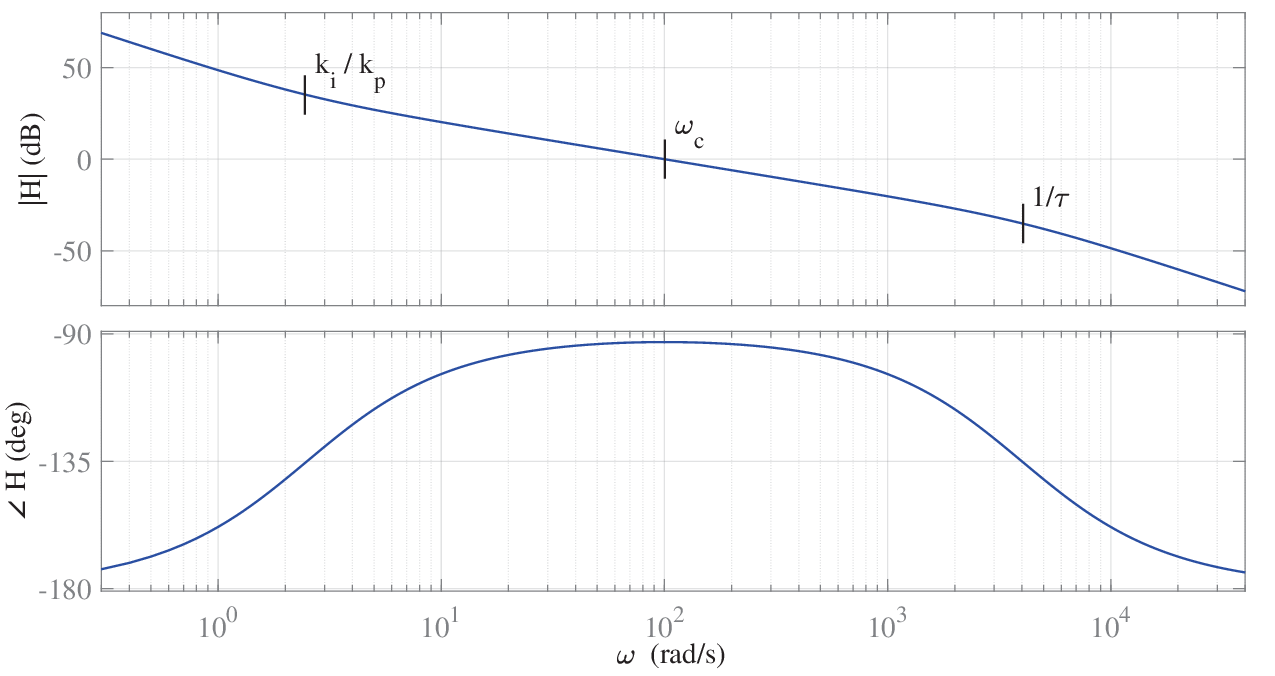}
\caption{Open-loop transfer function around crossover frequency $\omega_c$.}
\label{fig:symmoptimum}
\end{figure}

For tuning linear feedback controllers with a loop transfer function which has possibly flat amplitude characteristics around the crossover frequency $\omega_c$, the symmetrical optimum method, introduced by Kessler in \cite{kessler1958}, can be used for maximizing the value of the phase
margin, cf. \cite{voda1995,preitl1999}. Given the loop transfer function of the form \eqref{eq:2:7}, the normalizing factor $\alpha > 1$, cf. \cite{kaura1997}, can be used as the single design parameter which relates the crossover frequency to the PI control gains $k_p$ and $k_i$. For the smallest time constant in the loop, that is for $\tau$, the $\alpha$-factor scales down the crossover frequency relative to the corresponding largest corner frequency $1/\tau$, leading to
\begin{equation}
\omega_c = (\alpha \tau)^{-1}.
\label{eq:2:11}
\end{equation} 
The corresponding PI control gains, which provide a symmetrical optimum of both corner frequencies around the crossover one and for the plant gaining factor $U$, are then given by 
\begin{eqnarray}
\label{eq:2:12}
  k_p &=&  \frac{1}{U} \, \frac{1}{\alpha \tau},\\[1mm]
  k_i &=&  \frac{1}{U} \, \frac{1}{\alpha^3 \tau^2}. 
\label{eq:2:13}
\end{eqnarray}
Substituting $U=\sqrt{2/3}$ into \eqref{eq:2:12}, \eqref{eq:2:13}, this with respect to \eqref{eq:2:6} and \eqref{eq:2:7}, one obtains a simple tuning rule for the SRF-PLL. This will maximize the stability criteria in view of the sampling delay $\tau$ and the associated third-order denominator dynamics of the loop transfer function $H(s)$. A typical open-loop transfer function around the crossover frequency $\omega_c$ is exemplary shown in Fig. \ref{fig:symmoptimum}.
It is easy to recognize that since the $\bigl|H(j\omega)\bigr|$ curve proceeds with $-40$ dB/dec decrease beyond both corner frequencies $k_i/k_p$ and $1/\tau$ and with $-20$ dB/dec in between, the symmetrical optimum provides the maximum phase value at $\omega_c$. Worth noting is that the corresponding phase margin, which is defined as $\varphi_m = 180 + \angle H(j\omega_c)$, has the value $\varphi_m \rightarrow 90$ deg only for $\alpha \rightarrow \infty$. For the finite values of the $\alpha$-factor, however, the phase margin is $0 < \varphi_m < 90$ deg; cf. with Fig. \ref{fig:symmoptimum} for which $\alpha = 40$ is assumed.

\section{Robust frequency estimation}
\label{sec:3}

The robust frequency estimator \cite{ruderman2022} identifies online the angular frequency $\omega$ of a harmonic signal, i.e. $Z_{n}(t)$, $n \in \{a,b,c\}$, provided the latter is unbiased and the frequency variations $|d \omega / dt|$ are sufficiently slow comparing to the convergence rate of the estimation variable $\tilde{\omega}(t)$.      

The 2nd-order estimator dynamics is given by, cf. \cite{ruderman2022},
\begin{eqnarray}
\nonumber \left(%
\begin{array}{c}
  \dot{\eta}_1 \\
  \dot{\eta}_2 \\
\end{array}%
\right) & = & \left(%
\begin{array}{cc}
  0 & 1 \\
  -\tilde{\omega}^2 & -2 \tilde{\omega} \\
\end{array}%
\right) \left(%
\begin{array}{c}
  \eta_1 \\
  \eta_2 \\
\end{array}%
\right) + \left(%
\begin{array}{c}
  0 \\
  2 \tilde{\omega} \\
\end{array}%
\right) Z_n,
\\[1mm]
\nu & = & \left(%
\begin{array}{cc}
  0 & 1 \\
\end{array}%
\right) \left(%
\begin{array}{c}
  \eta_1 \\
  \eta_2 \\
\end{array}%
\right), \label{eq:3:1}
\end{eqnarray}
with the frequency adaptation law 
\begin{equation}
\dot{\tilde{\omega}} = -\gamma  \,
\mathrm{sign}(\eta_1)\bigl(Z_n - \nu\bigr). \label{eq:3:2}
\end{equation}
The only assignable parameter is the estimation gain factor $\gamma > 0$, while an initialization by $\tilde{\omega}(0) > \omega$ is recommended, cf. \cite{ruderman2022}. Recall that the robust estimation algorithm \cite{ruderman2022} was inspired by the globally convergent frequency estimator \cite{hsu1999}, while offering an improved performance and robustness to the perturbations comparing to \cite{hsu1999}. The number of the tuning parameters of the algorithm was also reduced to one. \textcolor[rgb]{0.00,0.00,0.00}{Practically, the choice of $\gamma$-value constitutes a tunable trade-off between the convergence speed (i.e. higher values imply faster convergence) and the application-specific signal-to-noise ratio. Higher $\gamma$-values will make estimation more noise sensitive and lead to higher spurious fluctuations in the dynamic $\tilde{\omega}(t)$ value.}  

Assuming a given input harmonic signal 
\begin{equation}
Z_n(t) = Z \sin \bigl(\omega t + \theta \bigr) + \mu(t), \label{eq:3:3}
\end{equation}
with injection of a band-limited and zero-mean noise $\mu$, the convergence rate of the robust frequency estimator \cite{ruderman2022} is provided by the following theorem.

\begin{theorem}
\label{thm:1} The frequency estimator \eqref{eq:3:1}-\eqref{eq:3:2} of \eqref{eq:3:3}
converges asymptotically as $\tilde{\omega}(t) \rightarrow \omega$ for
$t \rightarrow \infty$, provided the gain $\gamma > 0$ is appropriately tuned with respect to the noise signal $\mu$. The frequency estimation error $\varepsilon (t) = \tilde{\omega}(t) - \omega$ converges uniformly and exponentially in terms of
\begin{equation}
\bigl | \varepsilon(t_2) \bigr | < \alpha \bigl | \varepsilon(t_1)
\bigr | \exp \bigl( - \beta (t_2-t_1) \bigr), \label{eq:3:4}
\end{equation}
for some $\alpha > 0$ and $\forall \; t_2>t_1$. The exponential
rate of convergence is independent of $\mu$ and is upper bounded by
\begin{equation}
\beta = \frac{1}{2} \, \gamma Z \, \omega^{-1} + \delta, \label{eq:3:5}
\end{equation}
where $\delta$ is a small positive constant.
\end{theorem}

\begin{IEEEproof}
The proof of Theorem \ref{thm:1} is provided in \cite{ruderman2022}.
\end{IEEEproof}

\begin{remark}
While $\gamma$ and $Z$ are assumed to be constant for the designed SRF-PLL with feed-forward, cf. Fig. \ref{fig:PLL}, the $\omega$ value is a process variable and hence not assignable. At the same time, it is affecting the convergence rate \eqref{eq:3:5}. Thus, a conservative estimate $\max(\omega)$ has to be followed in the applications.
\end{remark}

An example of the frequency estimation of the phase current $n = a$ measured on the side of an inverter-fed PMSM (cf. section \ref{sec:4} below) is shown in Fig. \ref{fig:freest} for the sake of illustration. The estimated angular frequency, starting from the time $t=3$ sec, is depicted over the measured one in the plot (a). Here we notice that the measured frequency refers to the electrical rotor speed, i.e., the mechanical speed measured by the encoder multiplied by the number of the pole pairs. While this is not strictly identical to the electrical frequency of the stator current signal, especially under short-term transients or distortions, it serves as a meaningful reference under typical operating conditions. A more direct synchronization assessment, based on the waveform-level comparison, is also presented in Section \ref{sec:4}. The assigned adaptation gain and the initial value are $\gamma = 4000$ and $\tilde{\omega}(0)=120$ rad/s, respectively. Note that for $t<3$ sec, the measured input harmonic has a very low signal to noise ratio, this due to an unloaded PMSM operation, which makes the $Z_a(t)$ signal unusable for any online frequency estimation. Each time the electromagnetic torque of the PMSM was increased stepwise, both the measured $\omega(t)$ and the estimated $\tilde{\omega}(t)$ experience the transient peaks, followed by a settling phase, see Fig. \ref{fig:freest} (a).
\begin{figure}[!h]
\centering
\includegraphics[width=0.98\columnwidth]{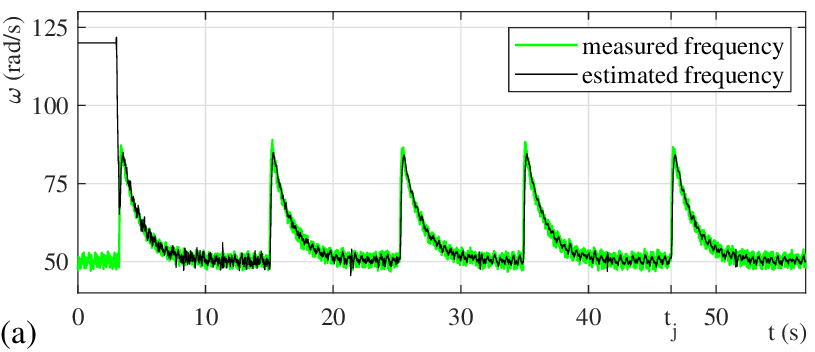}
\includegraphics[width=0.98\columnwidth]{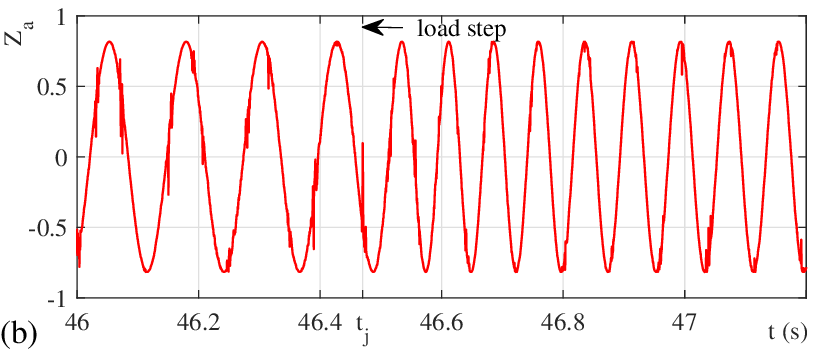}
\caption{(a) measured and estimated (starting from $t=3$ s) angular frequency for the assigned $\gamma = 4000$, $\tilde{\omega}(0)=120$ rad/s; (b) excerpt from the measured input at one of the time  instants $t_j$ of the load step.}
\label{fig:freest}
\end{figure}
An excerpt from the used $Z_a(t)$ measurement (with amplitude  normalized by means of \eqref{eq:2:5}) is exemplary depicted in Fig. \ref{fig:freest} (b) for one of the time instants $t_j$ of the torque step. The notching high-frequency peaks, which appear as inherent disturbances for any harmonic parameters estimation, are also visible in the measured $Z_a(t)$ pattern. 

\begin{figure}[!h]
\centering
\includegraphics[width=0.48\columnwidth]{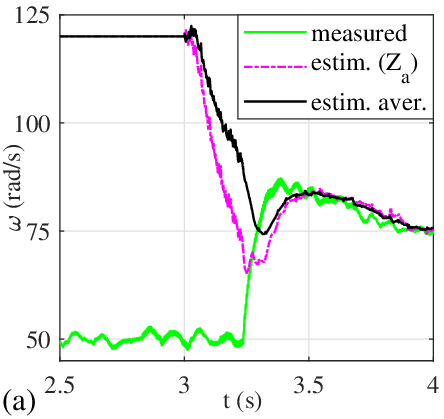}
\includegraphics[width=0.48\columnwidth]{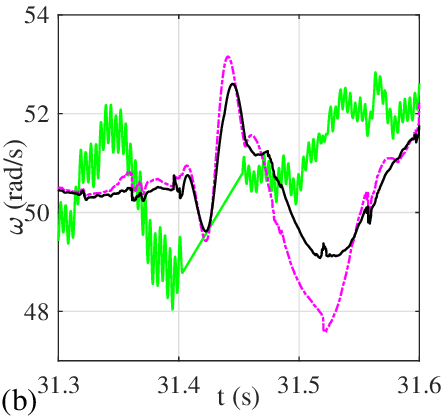}
\caption{Frequency tracking excerpts during the convergence (a) and temporary measurement lose (b) -- measured frequency versus single phase based estimation and averaged estimation from the three phases.}
\label{fig:freest3phases}
\end{figure}

In order to average the $\tilde{\omega}(t)$ estimate and, this way, further increase robustness and also improve the transient smoothness in case of a temporary lose of the measured input data, three equal estimators \eqref{eq:3:1}, \eqref{eq:3:2} are executed in parallel for $\bm{Z}_{\{a,b,c\}}$. The resulted frequency estimate, which is then entering the SRF-PLL in feedforwarding (cf. Fig. \ref{fig:PLL}), is 
\begin{equation}
\tilde{\omega}(t) = \frac{1}{3} \bigl( \tilde{\omega}_a(t)  + \tilde{\omega}_b(t) + \tilde{\omega}_c(t) \bigr). \label{eq:3:6}
\end{equation}
The achieved effect is visualized in Fig. \ref{fig:freest3phases} by using the same experimental data as shown in Fig. \ref{fig:freest}.
Both frequency estimates, one based on the single phase signal ($Z_a$ taken here exemplary) and another which is averaged according to \eqref{eq:3:6}, disclose largely a very similar performance. The improvement becomes visible in the shown tracking excerpts. In Fig. \ref{fig:freest3phases} (a), the convergence phase with an associated lagging after the transient overshoot is highlighted. In Fig. \ref{fig:freest3phases} (b), a temporary measurement loss (cf. the straight line segment in the measurement curve) and the resulting transient peaking in the estimation signals can be seen. 

It it also worth noting that the used frequency estimation algorithm is highly robust also to the additional disturbing sub-harmonics. The $\tilde{\omega}(t)$ estimate converges to the global minima in frequency domain, cf. \cite{ruderman2022} for details, and does not reman locked on the sub-harmonics. This is especially relevant in three-phase systems, for instance in the motor drive applications, where the 3rd harmonic disturbance, i.e. additive periodic signal to $Z_n$ with $3\omega$ angular frequency, can be persistently present. The estimation algorithm \cite{ruderman2022} demonstrates robust performance, even when the ratio of the amplitude of the 3rd harmonic to the 1st (the main estimated) \cite{ruderman2022} is about 0.2, which represents a relatively unfavorable case in terms of the quality of the measured $Z_n$ signal. For the sake of brevity, the results that additionally illustrate such robust behavior are not explicitly reported in this work.

\subsection*{Overall SRF-PLL-FF architecture}
\label{sec:3:x}
    
\begin{figure}[!h]
\centering
\includegraphics[width=0.98\columnwidth]{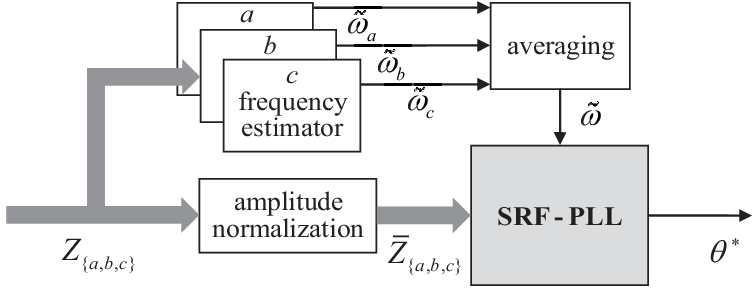}
\caption{Block diagram of the overall proposed SRF-PLL scheme with feedforwarding, cf. Fig. \ref{fig:PLL}.}
\label{fig:OverallScheme}
\end{figure}

The block diagram of the overall proposed SRF-PLL scheme with feedforwarding is shown in Fig. \ref{fig:OverallScheme}. The standard SRF-PLL, cf. Fig. \ref{fig:PLL}, is grey-shadowed, while the proposed extensions described above are depicted in the white transfer blocks. Note that three identical frequency estimators \eqref{eq:3:1}, \eqref{eq:3:2} are implemented, for three phases each, while the output of averaging \eqref{eq:3:6} is the resulted frequency estimate entering the SRF-PLL in feeforwarding, cf. Fig. \ref{fig:PLL}. Also recall that an amplitude normalization \eqref{eq:2:4}, \eqref{eq:2:5} is undertaken so as to render the PLL loop transfer characteristics insensitive to the amplitudes variation of the input harmonic signals.

Worth noting is that the complexity of the proposed SRF-PLL-FF scheme is not significantly higher compared to the standard SRF-PLL implementation. The white transfer blocks in Fig. \ref{fig:OverallScheme} requires solely $3 \times 3$ integrators in addition to several (less than 15) scalar algebraic expressions and auxiliary variables for a real-time implementation.

\section{Experimental study}
\label{sec:4}

\begin{figure}[!h]
\centering
\includegraphics[width=0.98\columnwidth]{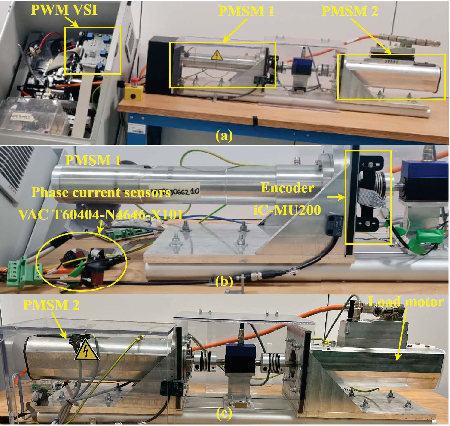}
\caption{Experimental setup (laboratory view).}
\label{fig:setup}
\end{figure}

In this section, we present a detailed experimental evaluation of the designed SRF-PLL with feed-forward estimation of the effective frequency $\tilde{\omega}$. A picture of the experimental setup (laboratory view) is shown in Fig. \ref{fig:setup}. The case study uses the three-phase currents of a voltage source inverter (VSI), which is utilizing the pulse-width modulated (PWM) insulated-gate bipolar transistor (IGBT) modules (Semikron SEMiX303GB12E4s) and operating at a switching frequency of 8kHz. The phase currents are measured using the Vacuumschmelze T60404-N4646-X101 current sensors during the field-oriented control (FOC) of two sensored permanent magnet synchronous motors (PMSMs). Each motor is equipped with a dedicated rotational encoder: an iC-MU200 magnetic absolute off-axis encoder for PMSM 1, and a Lika AMM8023/BG1-32-X1/S823 encoder for PMSM 2. The motion control logic is executed on an FPGA-based control platform (dSPACE 1006). The implemented frequency estimation algorithm and SRF-PLL loop are executed in the discrete-time numerical simulation (with the first-order Euler solver) while the sampling rate is set to the same value as the real-time sampled experimental data. 

The assigned estimator gain parameter is $\gamma = 4000$ for both, the torque-varying operation (cf. section \ref{sec:4:sub:2}) and frequency-ramp operation (cf. section \ref{sec:4:sub:3}). The PI gains are designed to be the same setting in both experiments by means of the symmetrical optimum approach (cf. section \ref{sec:2:sub:3}): $k_p = 122$ and $k_i=306$. \textcolor[rgb]{0.00,0.00,0.00}{Here we emphasize that the same PI control gains were used for both evaluated control schemes: the SRF-PLL without feedforwarding (i.e. $\tilde{\omega}$ set to zero), and the SRF-PLL-FF with $\tilde{\omega}$ estimate injected in feedforwarding, cf. Fig. \ref{fig:OverallScheme}.} The sampling time is $\tau = 0.00025$ sec, and no additional low-pass or other signal filters are used for acquisition of the experimental data. 

The SRF-PLL evaluation is performed by comparing the reference measured phase angle $\theta$ with the estimated one $\theta^{\ast}$. Moreover, the amplitude normalized $\bar{Z}_a(t)$ measurement and the correspondingly SRF-PLL computed
$$
\bar{Z}_a^{\ast}(t) = \bar{Z} \sin\bigl(\theta^{\ast}(t)\bigr) 
$$
are exemplary shown for the load-varying operation, this for a better traceability of synchronization of the harmonic signals.

\subsection{Torque-varying operation}
\label{sec:4:sub:2}

The first evaluation is carried out on the phase currents of PMSM 2 during the torque-varying operation. In particular, during these experiments, the PMSM 2 is torque controlled while the load motor (shown in Fig. \ref{fig:setup} (c)) is speed controlled.

Two experimental data sets are evaluated, one with a steady-state angular frequency of 50 rad/s (cf. Fig. \ref{fig:freest}) and another one with a steady-state angular frequency of 150 rad/s. Note that both experiments contain a series of stepwise torque changes, five times in total (cf. Fig. \ref{fig:freest} (a)), so that the magnitude variation of the sampled three-phase current is $Z \in [0.5,\ldots,1.5]$ A.     

\begin{figure}[!h]
\centering
\includegraphics[width=0.98\columnwidth]{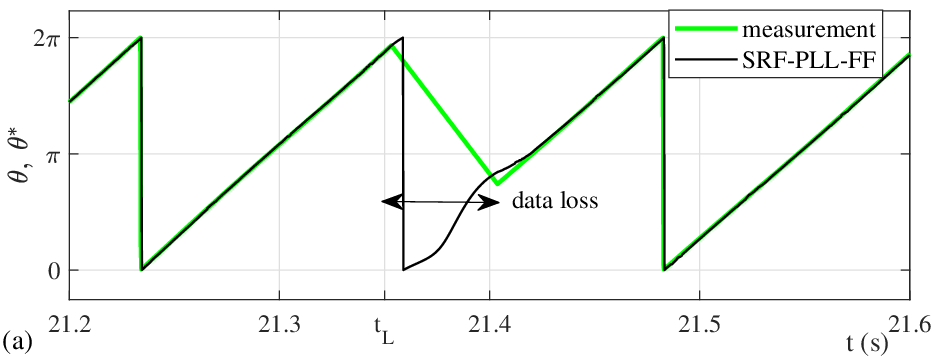}
\includegraphics[width=0.98\columnwidth]{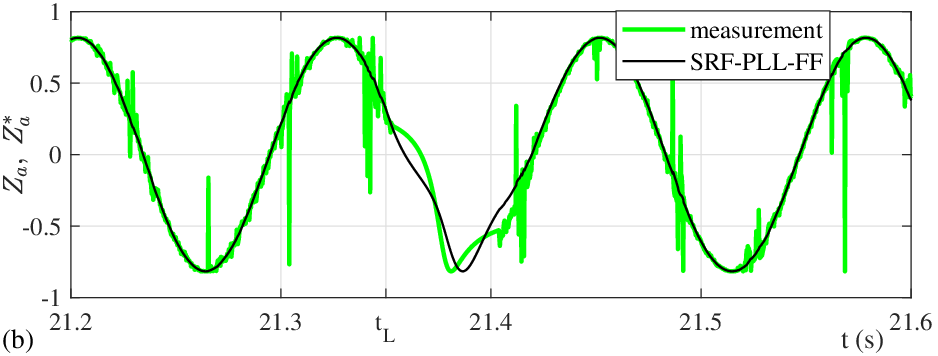}
\caption{Comparison of the measurement and SRF-PLL-FF estimation for 50 rad/s experiment, fragment with a transient data loss: phase angle $\theta$ (a) and amplitude normalized $\bar{Z}_a$ current signal (b).}
\label{fig:50estim}
\end{figure}

A time fragment which includes also a transient data loss (starting from the marked time instant $t_L$) is shown in Fig. \ref{fig:50estim} for the 50 rad/s experiment. Here, the feedforward frequency estimation is used with SRF-PLL (denoted by SRF-PLL-FF). Note that only three periods are depicted for the sake of a better visualization. In plot (a), the phase reference is derived from the electrical rotor angle measured via encoder. A more direct validation is provided by the comparison of the measured and reconstructed phase current waveforms in plot (b). An interval of the temporary missing data is visible in both (a) and (b) plots, while the relatively high notching disturbances of the measured $\bar{Z}_a(t)$ input signal are also visible in the plot (b).  

\begin{figure}[!h]
\centering
\includegraphics[width=0.98\columnwidth]{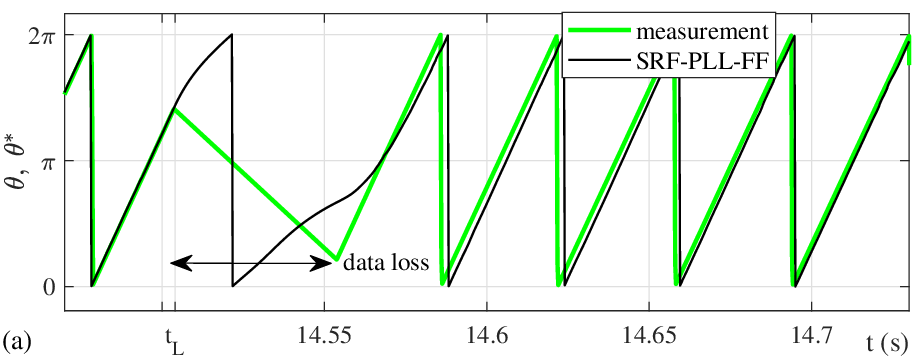}
\includegraphics[width=0.98\columnwidth]{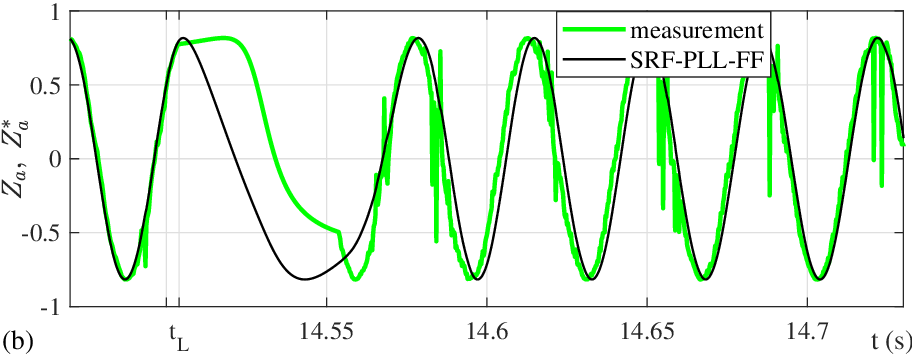}
\caption{Comparison of the measurement and SRF-PLL-FF estimation for 150 rad/s experiment, fragment with a transient data loss: phase angle $\theta$ (a) and amplitude normalized $\bar{Z}_a$ current signal (b).}
\label{fig:150estim}
\end{figure}

The similar SRF-PLL-FF performance can be observed also from the tripled frequency experiment of 150 rad/s, as fragmentary shown in Fig. \ref{fig:150estim}. Also here, a transient data loss (starting from the marked time instant $t_L$) is picked out, after which the synchronization is recovered  within just the few periods. Note that during the recovery time, the phase angle error is limited and continuously decreasing, while the phase lock remains operational also during the temporary data loss of about 0.05 s, that is essential comparing to $\tau$.

\subsection{Frequency ramp operation}
\label{sec:4:sub:3}

The second evaluation scenario examines the phase currents of the PMSM 1 during a controlled ramp start-up sequence. During the experiment, the motor is loaded by the PMSM 2 which is controlled at a constant torque, cf. Fig. \ref{fig:setup} (b).

\begin{figure}[!h]
\centering
\includegraphics[width=0.98\columnwidth]{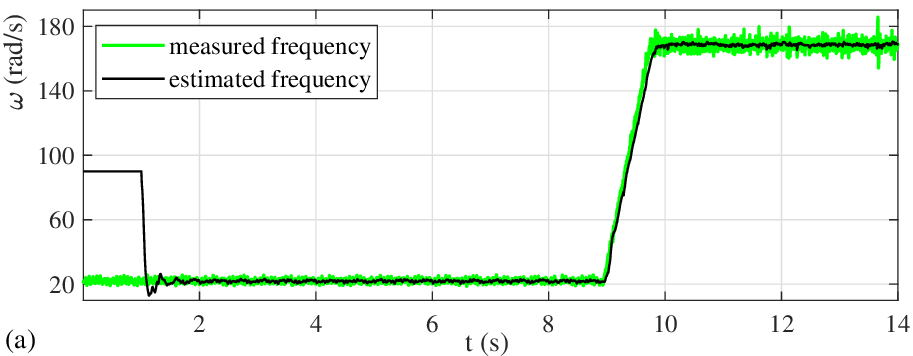}
\includegraphics[width=0.98\columnwidth]{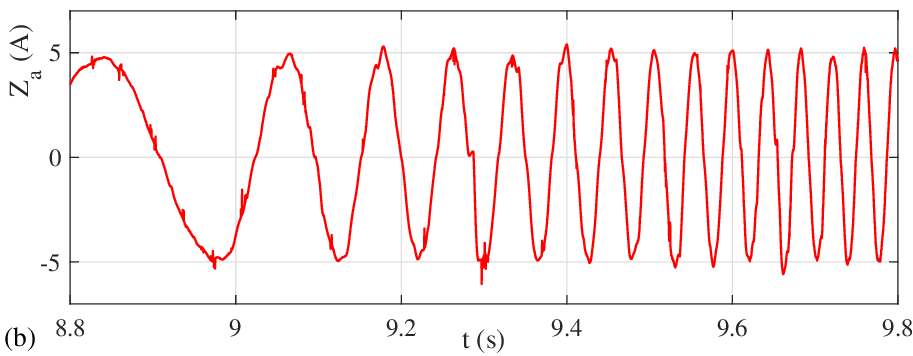}
\caption{(a) measured and estimated (starting from $t=1$ s) angular frequency for the assigned $\gamma = 4000$, $\tilde{\omega}(0)=90$ rad/s; (b) excerpt from the measured input current during the frequency ramp.}\label{fig:freqestramp}
\end{figure}

The measured and feedforward estimated angular frequencies $\omega(t)$ and $\tilde{\omega}(t)$, respectively, are shown opposite each other in Fig. \ref{fig:freqestramp} (a). The same encoder-based reference is used here, consistent with the approach discussed before. One can recognize that both coincide well with each other, also during a relatively fast ramp. Recall that the latter is essential for PMSM operation at large, during the acceleration and deceleration phases, cf. \cite{novak2022}. An excerpt from the measured input current during such frequency ramp is depicted in Fig. \ref{fig:freqestramp} (b) for better traceability.

\begin{figure}[!h]
\centering
\includegraphics[width=0.98\columnwidth]{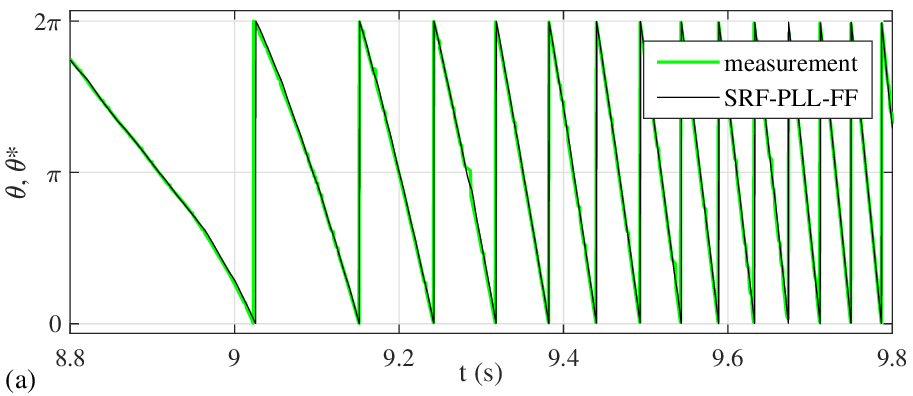}
\includegraphics[width=0.98\columnwidth]{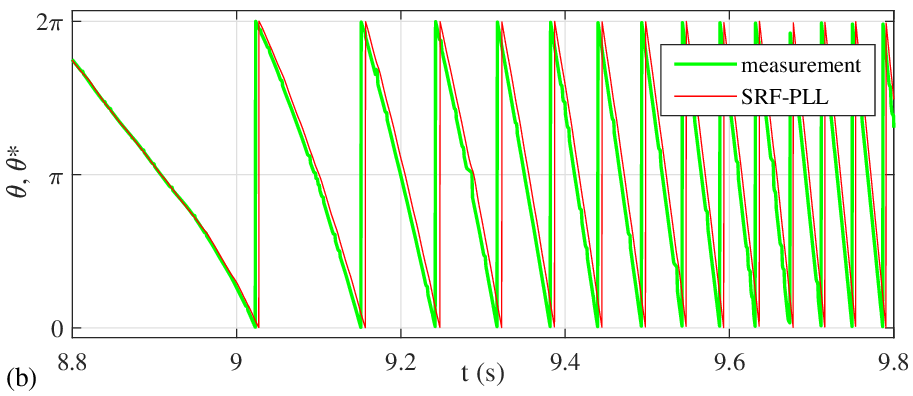}
\caption{Comparison of the measurement and PLL estimation during the frequency ramp: (a) phase angle $\theta$ for SRF-PLL with feed-forward (SRF-PLL-FF), (b) phase angle $\theta$ for SRF-PLL without feed-forward i.e. $\tilde{\omega}=0$.}\label{fig:PLLramp}
\end{figure}

A comparison of the measured and PLL-estimated phase angle during the frequency ramp is shown in Fig. \ref{fig:PLLramp}, once with the use of the feedforward frequency estimation in (a) and once without it in (b). In both cases the SRF-PLL is able to lock the phase angle, but the synchronization  progresses much faster in case of using the feedforward (SRF-PLL-FF), as it is visually recognizable from both diagrams.

\subsection{Evaluated error metrics}
\label{sec:4:sub:4}

The phase synchronization error $\Delta \theta$ is evaluated quantitatively for both the load-varying operation and frequency ramp operation described above. Two error metrics, the accumulative absolute error ($\Sigma$) and the absolute mean error (ME) 
\begin{equation}
E_{\Sigma} = \sum \limits_{k}^{K} \bigl| \Delta \theta(t_k) \bigr| \quad \hbox{ and } \quad E_{\textrm{ME}} = \frac{1}{K} \sum \limits_{k}^{K} \bigl| \Delta \theta(t_k) \bigr|,
\label{eq:4:1}
\end{equation} 
respectively, are calculated for the recorded real-time series with the sampling index $t_k$. Note that while for both load-varying operation experiments, i.e. with $50$ rad/s and $150$ rad/s steady-state frequencies, the entire time series of about $55$ sec are evaluated, cf. Fig.  \ref{fig:freest}, the frequency ramp operation is evaluated for $9 < t < 10$ sec only, where the frequency ramp is appearing, cf. Fig. \ref{fig:freqestramp} (a). The error metrics are listed in Table \ref{tab:1}
in comparison between the SRF-PLL and the same SRF-PLL augmented by the feedforward frequency estimation (here abbreviated by PLL-FF).
\begin{table}[!h]
  \renewcommand{\arraystretch}{1.5}
  \caption{Evaluated error metrics for the phase synchronization error}
  \label{tab:1}
  \footnotesize
  \begin{center}
  \begin{tabular} {|p{3.4cm}||p{2cm}|p{2cm}|}
  \hline
  Experiment  &  $E_{\Sigma}$  & $E_{\textrm{ME}}$  \\
  
              &   PLL / PLL-FF   &  PLL / PLL-FF \\
  \hline \hline
  Load-varying oper. $50$ rad/s        & $1.21e4$ / $7.82e3$    &  $0.0584$ / $0.0376$          \\
  \hline
  Load-varying oper. $150$ rad/s       & $2.44e4$ / $1.42e4$    &  $0.1185$ / $0.0716$          \\
  \hline
  Frequency ramp oper.                 & $3.68e3$ / $0.53e3$    &  $0.9205$ / $0.1327$       \\
  \hline
  \end{tabular}
  \end{center}
\end{table}
Further we note that due to the periodic jumps (i.e. reset) in both $\theta(t)$ and $\theta^{\ast}(t)$ quantities by reaching $2\pi$ rad and $0$ rad, cf. Figs. \ref{fig:50estim}, \ref{fig:150estim} and \ref{fig:PLLramp}, a spurious large peaking of $\Delta \theta$ is appearing at each full rotation of the phase angle. Therefore, the evaluated error metrics $E_{\Sigma}$ and $E_{\textrm{ME}}$ represent only the relative values used for the sake of comparison. However, both error metrics are justified by having the same PLL design, the same data sampling, and the same signal processing, so that the relative numbers given in Table \ref{tab:1} disclose a reasonable performance comparison between the standard SRF-PLL and SRF-PLL with feedforwarding.

Finally, to complete the assessment of the synchronization performance, another quantitative error metric based on the waveform reconstruction is also considered. Here, the root mean square (RMS) error between the measured and reconstructed current signals is computed as  
\begin{equation}  
E_{\text{RMS}} = \sqrt{ \frac{1}{K} \sum_{k=1}^{K} \left( \bar{Z}_a(t_k) - \bar{Z}_a^{\ast}(t_k) \right)^2}.
\label{eq:4:2}  
\end{equation}  
The RMSE values are calculated over the same intervals used for the phase error metrics. The results are summarized in Table \ref{tab:2}, highlighting a consistent reduction in the waveform error when feedforward estimation (PLL-FF) is applied, especially for the frequency ramp operation. On should note, however, that also the evaluated RMSE metric represents only a relative measure of the waveform reconstruction error. This is due to the relatively high notching disturbances of the measured input signal, cf. the plots (b) in Figs. \ref{fig:freest}, \ref{fig:50estim}, and \ref{fig:150estim}.
\begin{table}[!h]
  \renewcommand{\arraystretch}{1.5}
  \caption{Waveform-based RMSE of PLL-FF versus PLL}
  \label{tab:2}
  \footnotesize
  \begin{center}
  \begin{tabular} {|p{4cm}||p{3cm}|}
  \hline
  Experiment  &  $E_{\text{RMS}}$ \\ & PLL / PLL-FF \\
  \hline \hline
  Load-varying oper. $50$ rad/s   &  $0.0855$   /    $0.0777$       \\
  \hline
  Load-varying oper. $150$ rad/s  &  $0.1465$   /    $0.1000$      \\
  \hline
  Frequency ramp oper.         &     $0.2909$   /     $0.0513$      \\
  \hline
  \end{tabular}
  \end{center}
\end{table}

\section{Conclusion}
\label{sec:5}

This paper proposes a robust feedforward extension of the conventional three-phase synchronous reference frame phase-locked loop (SRF-PLL) by an online frequency estimator introduced in \cite{ruderman2022}. A small-signal control-oriented model of SRF-PLL is first recalled for analysis and design of the underlying PI feedback control part. We use a standard symmetric optimum approach \cite{kessler1958} for maximizing the phase margin at the crossover frequency of the PLL loop transfer function, and deliver a one-parameter robust PLL tuning in a manner originally provided in \cite{kaura1997}. Our SRF-PLL contribution is also in the shown power-invariant normalization scheme which renders the PLL loop gaining characteristics to be independent of the input amplitude, cf. \cite{kaura1997,golestan2017}. The feedforward extended PLL (SRF-PLL-FF) is entirely implemented to be real-time executable and to run without any additional filters and signals pre- or post-processing. The experimental case that we show, is on the 3-phase harmonic currents from the standard PMSM drives which are feeded by a voltage source inverter operating at the switching frequency of 8 kHz. The collected and used real-time data are sampled at 4 kHz. By evaluation the synchronized phase angle, we demonstrate that the SRF-PLL-FF performs clearly superior in comparison with the same SRF-PLL but without feedforwarding. Two steady-state frequency experiments, with 50 rad/s and 150 rad/s, both with stepwise increases of the load and thus transient peaking of the angular frequency are examined. In addition, a frequency ramp experiment that is typical for PLLs in context of the controlled electric motors is executed and assessed. It is believed, that the proposed feedforward SRF-PLL extension constitutes a considerable possible improvement (at zero hardware costs) for the SRF-PLL techniques which are widely used in the motor drives and power electronics at large.

\section*{Acknowledgments}
The first author acknowledges the financial support by NEST (Network for Energy Sustainable Transition) foundation during the sabbatical year at Polytechnic University of Bari.

\bibliographystyle{IEEEtran}
\bibliography{references}


 





\end{document}